# Surface Plasmon Polariton Assisted Optical Switching in Noble Metal Nanoparticle Systems: A Sub-Band Gap Approach

Sandip Dhara

Surface and Nanoscience Division, Materials Science Group, Indira Gandhi Centre for Atomic Research, Kalpakkam-603 102, India

Email : dhara@igcar.gov.in

## 1. Introduction

Understanding the light-matter interaction at nanometre scale is a fundamental issue in optoelectronics and nanophotonics which are prerequisites for advanced sensor applications, namely, optical switching (abrupt change in resistance with exposed light signal) devises using all-optical signal processing [1,2], two photon absorption (TPA) enhanced second harmonic generation (SHG) [3], renewable energy resourcing by guiding and localizing light and considerable reduction in absorption layer thickness [4], and in the bio-organic sensing and in medical therapy encompassing early detection of beginning of life using shift in plasmonic resonance frequency in the presence motion of molecular motors [5] and in controlling tumour by generating heat using the adiabatic heating of plasmonic vibration [6].

   Optical switching using the enhanced third-order nonlinear susceptibility in noble metal systems [7-10], especially near the surface-plasmon-resonance (SPR) frequency evoked considerable interest among scientific researchers during last one and half decade.  Coherent oscillation of conduction electrons, particularly in noble metal nanoclusters, with the excitation of visible light gives rise to surface plasmon polaritons (SPP) which propagate near the metal-dielectric interface (schematically shown in Fig. 1) [11-14]. The evanescent field allows the observation of interference [15-17], and sub-diffraction limited optical imaging of nanostructure [18-22]. The plasmon coupling within arrays of metal nanoparticles can lead to the formation of nanoscale hot spots in which the intensity of light from an incident beam can be concentrated by more than four orders of magnitude. The effect of light concentration by means of plasmon is most obvious in phenomena dealing nonlinearity in light intensity, as demonstrated recently by the on-chip generation of extreme-ultraviolet light by pulsed laser high harmonic generation [23]. This opens an affluence of prospects in lithography or imaging at the nanoscale through the use of soft x-rays. In fact, SPR induced localized SHG using Au nanoparticles are well studied system [24-26]. Among many such significant phenomena and applications, the most well known is the giant surface-enhanced Raman scattering (SERS) [27] or noble metal coated tip enhanced Raman scattering (TERS) [28] those allow both detection, and spectroscopic imaging of a single molecule [29]. Electromagnetic energy transfer in chains of closely spaced metal nanoparticles was reported in the sub-diffraction limit by means of coupled plasmon modes [30,31]. In a dispersion model for coupled plasmon modes considering equi-spaced metal nanoclusters is developed using an analytical model that describes the near-field electromagnetic interaction between the particles in the dipole limit [30]. Coherent propagation with group velocities exceeding 0.1 $c$ was calculated in straight wires of dimension less than 0.1$\lambda$ (wavelength) and around sharp corners with bending radius less than wavelength of visible light. In another report using generalized Mie theory, the light-transport properties of Ag particles of 50-nm diameter, an optimum guiding conditions for an inter-particle spacing of 25 nm, and a corresponding 1/*e* signal-damping length of 900 nm was estimated [31]. These models of optical energy transport in a plasmonic chain are useful for sub-wavelength transmission lines within integrated optics circuits and for near-field optical microscopy in futuristic ultra fast photonic device applications.

Coinage elements of Au, Ag, Cu are extensively studied for plasmonic applications [32]. Ultrafast switching of 360 fs in Ag nanoclusters embedded in $SiO_2$ matrix, originating from self-diffraction of a pump pulse due to transient grating, was recorded in a femtosecond optical-Kerr-shutter experiment [7]. A very high optical nonlinearlity $\chi^{(3)} \sim 10^{-7}$ esu which was comparable [8] or one order lower[9] than that for Au nanoclusters system, was reported. An one order lower value of $\chi^{(3)} \sim 10^{-8}$ esu was recorded on Cu nanoclusters dispersed in silica matrix with picosecond nonlinear optical response [10]. In a microreactor approach [33], Au NPs embedded in silica matrix showed optical switching with the exposure of 532 nm, which was close to the SPR absorption peak for Au NPs. In the similar approach, an optical switching was also reported for Au functionalized GaN (Au-GaN) nanowires with an excitation of 532 nm [34]. Role of band conduction in GaN is ruled out in the absence of optical switching using 325 nm (~3.8 eV) excitation (energy being higher than the band gap of GaN ~ 3.4 eV). The SPR, on the other hand, in bimetallic nanocluster of noble metals of Au and Ag is also useful for various applications with the ability of tuning the absorption peak position [35-37]. The SHG in the Pt/Cu [38] and Ag/Cu [39] systems drew lot of attention. In a recent report photoresponse of bimetallic Au-Ag nanoparticle embedded soda glass (Au-Ag@SG) substrate was reported [40] for surface plasmon assisted optical switching using 808 nm excitation, which was away from the characteristic SPR peaks of Ag (~400 nm) and Au (~550 nm) nanoparticles. The observation suggested the possible role of TPA owing to the presence of interacting electric dipole in these systems.

In order to understand the electrical transport process in noble metal-insulator nanocomposite or dispersed noble metal nanocluster on dielectric matrix, various models are discussed in the literature. Banking on high $\chi^{(3)}$ values, optical switching was reported in the percolation threshold of Au nanoclusters in $SiO_2$ matrix [9], and Cu nanoclusters on $Si_3N_4$ film [41]. A reversible electronic threshold switching was reported in percolative Ag nanoclusters embedded in polymer matrix [42]. The optical switching originating from the excitation of the surface plasmon was recorded for metal–oxide–metal tunnelling junctions [43]. The SPP in the form of drifting hot electrons across the oxide barrier and tunnelling to the counter electrode in the evanescent field of SPR was made responsible for electrical transport mechanism. In the specially designed experiment [34], optical switching in Au-GaN nanowires with a sub-band gap excitation of 532 nm suggested possible role SPP assisted transport of electron in the system. The same mechanism was proposed for the propagation of electronic carrier belonging to the conduction electron of noble bimetals in the Au-Ag@SG system in understanding the observed photoresponse [40]. Contributions of interband and intraband transitions in noble metals and inter-particle separation were discussed in understanding the plasmonic coupling mechanism.

These models, for the first time, are discussed in ambit of having a sub-band gap feature in the SPP assisted photoresponse where transport of electrical carriers may manifest either at the percolation threshold with enhanced electro-magnetic field, and in the form of tunnelling current through the potential barrier at the Fermi level or in the propagation of plasmon coupled electrons at SPR for metal-dielectric composites.

## 1.1 A Percolative Pathway for Electrical Transport

Optical absorption and electrical transport properties of noble metal-dielectric nanocomposites can be understood from the theoretical investigation in the calculation of the dielectric function for a heterogeneous composite medium [44-46]. The Maxwell-Garnett theory (MGT), dealing predictions related to the existence of the optical dielectric anomaly observed in granular metal films (presently understood as absorption peak due to SPR of the noble metal clusters), was modeled for the calculation of optical properties [44]. However, MGT was limited in predicting observed percolation threshold in granular metals for the volume fraction of the metallic phase comparable to that of the matrix phase. On the other hand, the effective medium theory (EMT) was used for the calculation of the dielectric constant for composites, as well as predicted a percolation threshold for electrical conductivity [45]. However, unlike the MGT, dielectric anomaly could not be inferred in the EMT. Moreover, the predicted value of the percolation threshold was lower than that compared with the experimental reports. Later, based on a phenomenological model considering distribution of conducting and insulating phases, both the optical dielectric anomaly and the percolation threshold was developed for the unified understanding of the optical absorption and percolation transport properties of granular composite media [46]. At a specific composition with the fraction of conducting phase $p > 0.35$ a percolative pathway for electrical conduction was correctly estimated for Au-SiO$_2$ composite (Fig. 2). At the same time, as observed in experiments for the SPR absorption of Au clusters embedded in dielectric matrices was well described under the same formalism in the limit of the percolative compositions $0.1 < p < 0.8$ (Fig. 3). The percolating threshold was not predictable under MGT with optical anomaly existing even for $p=1$. Transition in infrared transmission for $p > 0.7$ was another success of the model where also MGT failed to estimate. Subsequently, a hopping model in a percolative pathway, based on the microstructure of composites, was developed where the effect of charging energy $E_c$ as a function of the conductance of paths linking grains of separation 's' and diameter 'd' were considered with the assumption of constant $s/d$ throughout the specimen [47]. Deviating from the phenomenological description, a band model was proposed in the percolating clusters [48], independent of microstructure, for estimating the activation energy from the intra-grain energy level splitting due to the finite size of the grains, apart from the usual electrostatic charging energy, $E_c$. Later on drawing an analogy with Efros- Shklovskii model [49], the role of 'site energy' was conceptualized in introducing correlated Coulomb gap [50]. A detailed study on electrical conductivity in the Ag nanoparticle embedded in glass matrix reported inter-grain hopping of charge carrier [51]. The characteristic temperature of the conductivity, $T_0=4\chi s E_c/k_B$, where $k_B$ is the Boltzmann's constant and $\chi=(2m^*\varphi)^{1/2}/\hbar$ is the effective inverse tunneling width. Here, $m^*$ is the effective mass of the charge carrier, and $\varphi$ is the energy barrier over which the charge must hop, $\hbar$ is the reduced Planck's constant. The expression for $E_c=2q^2s/[\kappa d^2(½+s/d)]$, where $\kappa$ is the permittivity of the insulating phase and $q$ is the electronic charge. The values $T_0$ was obtained from the slope of the temperature-dependent resistivity plot (Fig. 4a). In the scope of the sub-band gap hopping model, the trend in inter-particle separation with increased density of metallic phase, as estimated from the SPR peak analysis (Fig. 4b), was correctly calculated from the expression of $T_0$ and $E_c$ with $\varphi \sim 4$ eV for glass matrix and $d$, the average size of the Ag clusters $\sim 10$ nm. Thus the role of sub-band gap states in the percolating noble metal clusters embedded in the dielectric matrices can be understood for the electrical transport in achieving optical switching.

Electrical conductivity at a threshold noble metal composition, $p>0.6$ of percolative Au-SiO$_2$ composite was reported to increase abruptly along with a distinct change in the absorption pattern with infrared transmission increasing above the same threshold value. A collective effect of local field enhancement of the Au nanoclusters in combination with the increased light amplitude in resonant cavities formed between the surfaces of the optimally dense Au clusters embedded in the $SiO_2$ matrix in the percolation limit was made responsible for the experimental observations [9]. In a unique study, light induced electrical conduction was reported in randomly-distributed Cu nanoparticles with varying surface coverage on an optically transparent $Si_3N_4$ [40]. At a percolation threshold of 64% metal coverage, an optical switching was recorded corresponding to the peak wavelengths of the peak SPR (Fig. 5). Finite difference time domain (FDTD) simulation of the dielectric constants based on Drude-Lorentz-Sommerfeld model [52] in case of the Cu nanoparticles/$Si_3N_4$ system showed (Fig. 6) field enhancement at the percolation threshold. Thus the report demonstrated that Cu nanoparticle-embedded device can detect the SPR by simply monitoring the current. The value of $\chi^{(3)}$ was found to depend on noble metal content in Ag:$BiO_2$ [53] and Cu:$Al_2O_3$ [54] composites with highest values achieved at the percolation threshold. The strong $\chi^{(3)}$ is responsible for optical switching in metal composite systems.

## *1.2 A Tunnelling Route to Electrical Transport*

A reversible electronic switching effect was observed in Ag nanoparticles embedded in polymer films. A sharp change of up to six orders of magnitude in the current–voltage behavior are highly reversible for these nanocomposite materials, and are defined as threshold switching at a percolation threshold, 0.78 of the metallic coverage (Fig. 7) [42]. At the percolation threshold microstructure is characterized by near continuous chain of particles, with no conductive metallic path formed between the electrodes. At a specific gap of 2 nm or less between the particles, electric field of ~ $10^7$ V/cm was estimated in between the Ag particles for an applied voltage of 1

V between electrodes. A tunneling of electronic charges with relatively high current densities would occur through the potential barrier at the Fermi level in the field emission process at such a high field.

Narrow band photoresponsivities up to 60 mV /W into a l00 Ω impedance at 632.8 nm was observed in Ag-Al$_2$O$_3$-Al and Al-Al$_2$O$_3$-Al metal-insulator oxide-metal (MOM) devices [43]. The wavelength sensitivity of the detector could easily be varied over a wide range by coating the top electrode with different dielectrics, the upper limit being the SPP frequency of the top electrode. Optimum metal grating parameters lead to high efficient SPP excitation. The data showed significantly improved photon assisted tunneling (SPP photo-signal to background ratio) with an applied bias. Inside the metals, the SPP was assumed to decay partly into single particle excitations of electron which could drift to the oxide barrier and tunnel to the counter electrode. The photoresponse was thus determined by the generation of the electrons in metal films at the evanescent field of SPR and by the tunneling rates through the barrier. A mean free path of the hot electrons was estimated to be ~ 25 nm for the MOM junction. In an interesting report, detection of visible light by two dimensional (2D) Ag-Al$_2$O$_3$-Al based MOM tunnel junctions on glass with periodic array of bumps was reported [55]. The SPP at the periodic gratings decayed into single particle excitations, which had sufficient energy to tunnel through the oxide barrier. They obtained a rectification of the photoresponse at 476.9 nm due to the nonlinear *I-V* characteristic of the tunnel junction. In a fresh approach, a strong wavelength-dependent and reversible photoresponse was reported in a two-terminal device using a self-assembled ensemble of Au nanopeapodded silica nanowires under light illumination, whereas no photoresponse was observed for the plain silica nanowires (Fig. 8). The switching wavelength was found to match the SPR absorption leading to the generation of electrons in the evanescent field where the photogenerated electrons tunnel through the oxide nanowires owing to the propagation of SPP. The propagating hot electron in the 1D nanoscale wave guide was observed to travel an inter-particle separation of 100 nm which was one order higher than that reported for MOM structures [43].

## *1.3 A Propagative Surface Plasmon for Electrical Transport*

Albeit the origin of percolative pathway or tunneling of electrons at percolation threshold in explaining the observed photocurrent in noble metal-dielectric matrices, a plasmon coupling based propagation of SPP can always be envisaged in the understanding of optical switching. It is particularly important when pure quantum mechanical electron tunneling cannot be used for inter-particle separations above 10 nm. Thus the conception of SPP, where plasmonic coupling played a vital role in the formation of hot electron in evanescent field and its propagation, was the principal driving mechanism for the observed optical switching in noble metal-dielectric matrices. The role of sub-band gap states, however was clearly understood for both percolative model considering 'site energy' correlated to Coulomb gap in the granular system [48,50] and in tunneling model by definition with electron travelling through the potential barrier at the Fermi level.

In a recent study, optical switching for 1D Au-GaN nanowires and in a single nanowire (Fig. 9) was demonstrated [34] for sub-band gap (532 nm; GaN band gap at 365 nm) excitation with surface plasmon resonance peak of Au around 550 nm. Conduction process, responsible for the observed photoresponse (Fig. 10), was conceived as SPP assisted transport of electron with propagative resonating electromagnetic radiation coupled to Au nanoclusters generating the carriers (schematic in Fig. 11) [56]. Resistivity measurements in the single nanowire showed lowering of resistivity from the bulk value of 0.3 Ω-cm (measured at dark) to 0.05 Ω-cm for the 532 nm illumination as a measure of surface plasmon polariton assisted electrical conduction process [34].

In a novel study, optical switching in the Au-Ag@SG system using 808 nm excitation was reported [40]. A change in current was almost twice for 808 nm excitation than that observed for the exposure with 532 nm (Fig. 12) indicating a possible role of TPA in the conduction process. The effect of TPA, unlike the conventional wisdom, was realized in its manifestation of doubling the photocurrent with the excitation wavelength of 808 nm as compared to 532 nm excitation where a single photon was involved in the SPR assisted photoresponse. In the sub-band gap picture, the SPP assisted generation of photocurrent does not occur across the band so a quadratic increase in the current cannot be expected. The photocurrent was because of the transport of conduction electrons belonging to Au-Ag system, so the role of TPA was limited to the plasmonic coupling the conduction electrons of Au-Ag system. In order to understand the possible role of competitive inter- and intraband transition in Au and Ag, FDTD calculations were analyzed showing very strong dipole coupling of Au-Ag@SG system for both 532 nm and 808 nm excitations for an inter-particle separation between Au and Ag nanoparticles of 2 nm (Fig. 13a). On the other hand, absence of electromagnetic coupling for 405 nm for the same system was noteworthy. The observation was correlated to the absence of interband contribution of Ag [56] to the electro-magnetic light absorption process in the bimetallic system of Au and Ag. Dipole interaction between Au and Ag nanoparticles was reported active up to a gap of 4 nm for both 532 nm and 808 nm excitations (Fig. 13b). The role of larger layer being involved in the electrical conduction process at 808 nm (with low absorption of 808 nm in the bimetallic system) excitation and increasing the photocurrent to almost twice the value of that observed in for 532 nm exposure was ruled out in the presence of strong coupling of electromagnetic wave of both 532 and 808 nm in the Au-Ag system (Fig. 13). The observation suggested possibility of optical switching in noble bimetallic nanocluster system where long wavelength with higher skin depth can be used for the communication purpose [40].

**Figure Captions :**

**Fig. 1.** (a) Incident-light wavelength dependencies of the electrical field intensity for different slit periods and (b) electrical field intensity distribution in a nano-slit grating for 1500 nm excitation. (Ref. [14]@2013; Copyright©*IEEE; Applied for permission*)

**Fig. 2.** Normalized conductivity $\bar{\sigma}$ as a function of $p$ for Au-SiO$_2$ cermets. Data from Ref. [43]. Dashed line denotes EMT result. (Ref. [46]@1980,Copyright©*American Physical Society; Applied for permission*)

**Fig. 3.** Optical transmission as a function of light wavelength for a series of Au-SiO$_2$ composites. Data are from Ref. [43]. For clarity, the curves are displaced with respect to one another. The theoretical curves are normalized to the experimental values at 0.3 μm. The theoretical values of $p$ are labelled to the right of pairs of curves, whereas the experimental values of $p$ and the film thickness are given above each pair of curves. (Ref. [46]@1980,Copyright©*American Physical Society; Applied for permission*)

**Fig. 4.** a) Plot of ln $R$ versus $T^{-\frac{1}{2}}$ plot at various fluences. The inset shows resistance ($R$) versus temperature ($T$) plot. b) Optical absorption spectra of ion exchanged and irradiated samples at various fluences. (Ref. [51]@2001,Copyright©*Elsevier Science B.V.; Applied for permission*)

**Fig. 5.** (a) LSPR measurement results from samples with a surface coverage, $f_a$ of 55%, 64%, and 80%; each samples has a different resonance peak (55%: 706 nm; 64%: 743 nm; 80%: 789 nm). (b) I-V characteristics of samples with a surface coverage of 55%, 64%, and 80%; the plot of $f_a$= 64% shows a jump of about one order of magnitude (for the percolation threshold). (Ref. [41]@2010,Copyright©*American Optical Society; Applied for permission*)

**Fig. 6.** Calculated electric field intensity enhancement in the plane of the deposited nanoparticles (64% coverage sample). The local field intensity enhancement is depicted in the TEM image

(inset) using the linear color bar. (Ref. [41]@2010,Copyright©*American Optical Society; Applied for permission*)

**Fig. 7.** Left: $I-V$ characteristics of a plasma polymer film containing Ag nanoparticles (AgPPF) with different area filling factors $f_a$. Top left: Sample setup: Coplanar electrode arrangement (slit dimensions 500 μm×6 nm); placement of TEM grids is matched to film positions (slits) for the electrical measurements. Right: TEM micrographs of different nanostructural types (dark: metal; light: plasma polymer). (Ref. [42]@2003,Copyright©*American Institute of Physics; Applied for permission*)

**Fig. 8.** Photoresponse measurements. The room-temperature resistance response as a function of time to light illumination for plain silica nanowires (upper part) and gold nanopeapodded silica nanowires (lower part). Shaded (pink, excitation wavelength $\lambda_{ex}$ =635 nm; green, $\lambda_{ex}$ =532 nm; purple, $\lambda_{ex}$ =405 nm) and unshaded regions mark the light-on and light-off periods, respectively. (Ref. [33]@2006,Copyright©*Nature Publication Group; Applied for permission*)

**Fig. 9.** Focused ion beam assisted Pt contact electrodes in (a) ensemble (b) single GaN nanowire system. (Ref. [34]@2014,Copyright©*Springer; Applied for permission*)

**Fig. 10.** Photoresponse in a) ensemble b) single GaN nanowire samples with periodical dark and 532 nm illumination conditions. (Ref. [34]@2014,Copyright©*Springer; Applied for permission*)

**Fig. 11.** Propagation of surface plasmons on gratings. (a) There are three forms of surface plasmon polaritons (SPPs) around each gold pitch: *a*, SPPs oscillating at the top surface of gratings; *b*, SPPs oscillating though the slits and; *c*, SPPs oscillating at the bottom surface of gratings at the Ti–Si Schottky interface. (b) Plasmonic heat absorption calculated by FDTD (shown with a logarithmic scale for clarity). Most of the hot electron generation occurs at the bottom surface of the gold layer. ([Ref. [55]@2013, Copyright©*Nature Publication Group; Applied for permission*)

**Fig. 12.** Typical photoresponse of Au-Ag embedded in soda glass sample grown with $Au^+$ implantation at a fluence of $5 \times 10^{16}$ ions.cm$^{-2}$ with periodical dark and illumination to different laser wavelengths. Inset shows similar photoresponse of samples grown with $Au^+$ implantation at fluences of $2 \times 10^{16}$ and $3 \times 10^{16}$ ions.cm$^{-2}$. The studies show a double amount of change in current for 808 nm excitation than that observed for 532 nm exposure. (Ref. [40]@2015,Copyright©*American Institute of Physics; Applied for permission*)

**Fig. 13.** Comparison of the electric field intensity enhancement contours for the interaction of electromagnetic radiation of different wavelengths with 5 nm Au and Ag nanoparticles separated by a) 2 nm and b) 4 nm in a medium with refractive index *n*=1.52. Here *k* is the electro-magnetic wave propagation vector, *E* is the electrical field vector and intensity bars indicate $|E|^2$. (Ref. [40]@2015,Copyright©*American Institute of Physics; Applied for permission*)

**Figures with Captions**

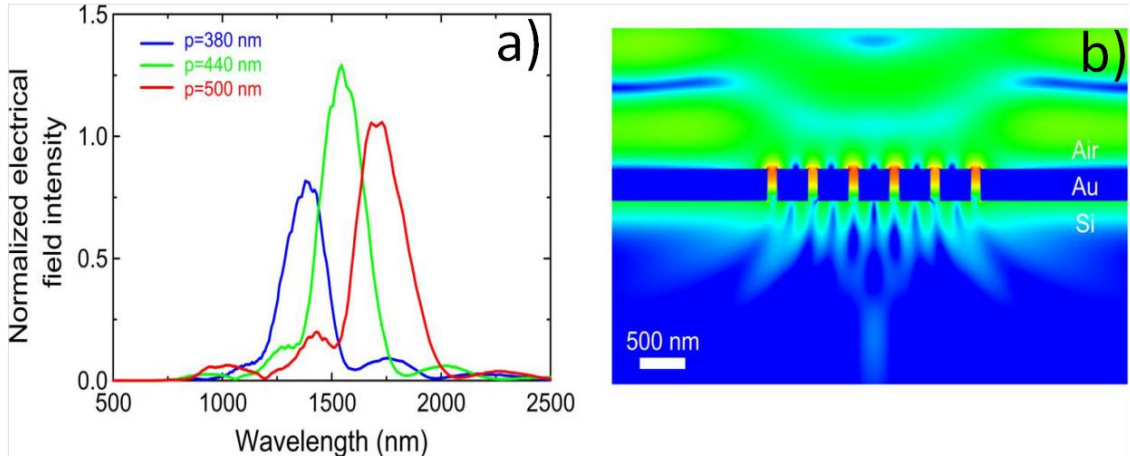

**Fig. 1.** (a) Incident-light wavelength dependencies of the electrical field intensity for different slit periods and (b) electrical field intensity distribution in a nano-slit grating for 1500 nm excitation. (Ref. [14]@2013; Copyright©*IEEE; Applied for permission*)

Page **22** of **34**

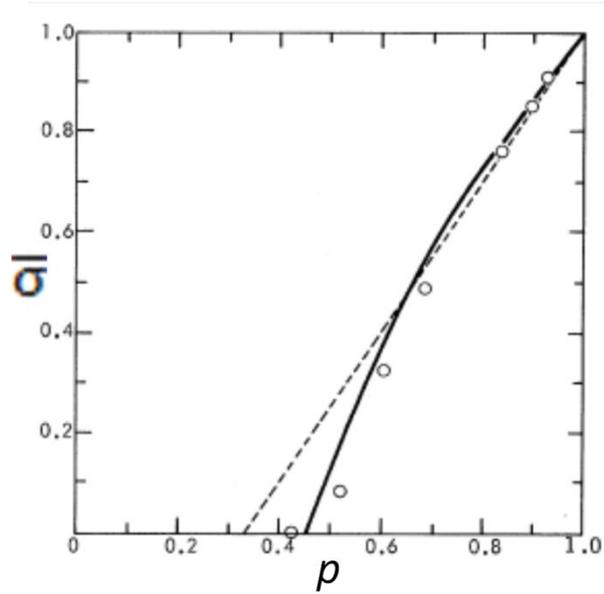

**Fig. 2.** Normalized conductivity σ̄ as a function of *p* for Au-SiO$_2$ cermets. Data from Ref. [43]. Dashed line denotes EMT result. (Ref. [46]@1980,Copyright©*American Physical Society; Applied for permission*)

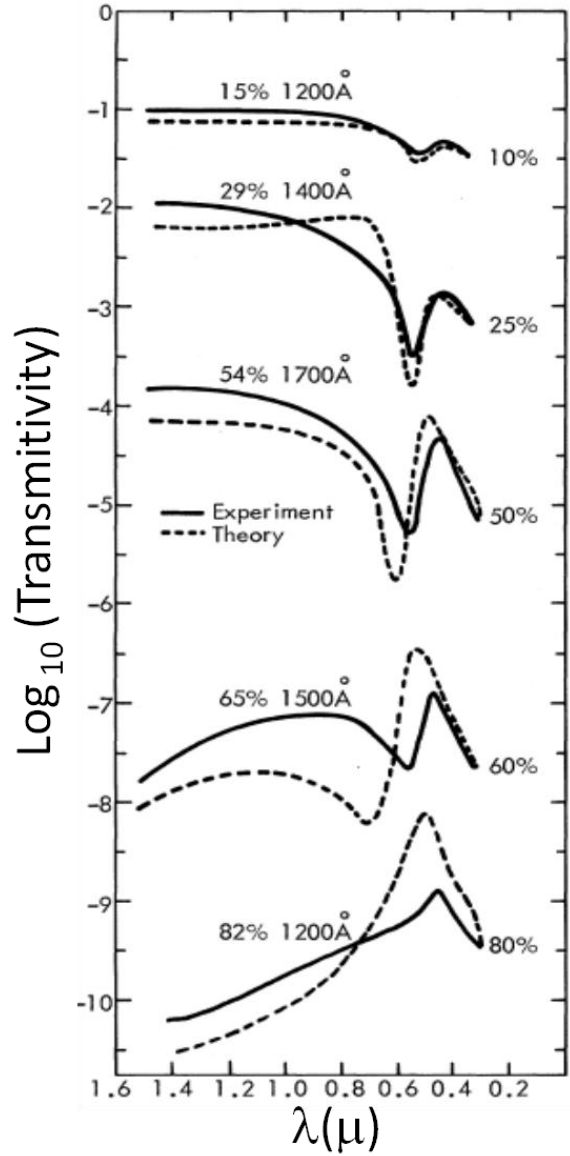

**Fig. 3.** Optical transmission as a function of light wavelength for a series of Au-SiO$_2$ composites. Data are from Ref. [43]. For clarity, the curves are displaced with respect to one another. The theoretical curves are normalized to the experimental values at 0.3 μm. The theoretical values of *p* are labeled to the right of pairs of curves, whereas the experimental values of *p* and the film thickness are given above each pair of curves. (Ref. [46]@1980,Copyright©*American Physical Society; Applied for permission*)

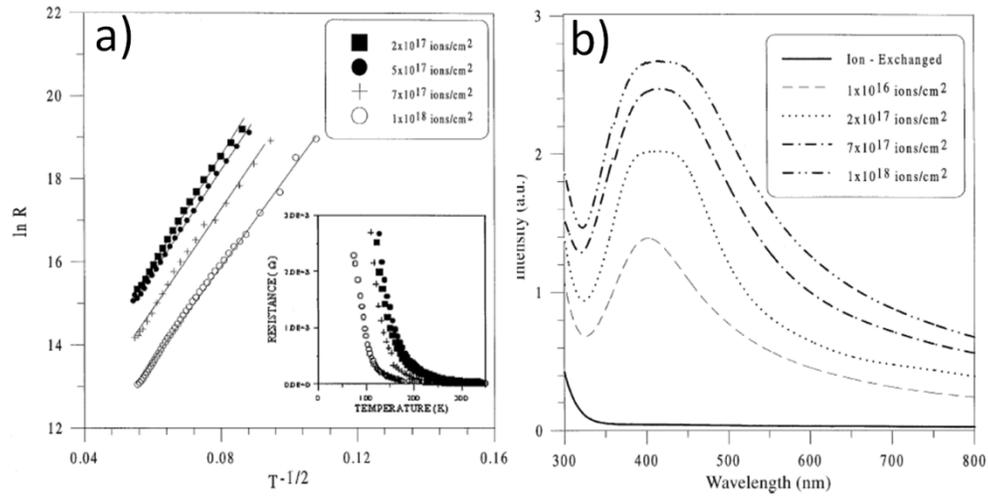

**Fig. 4.** a) Plot of ln $R$ versus $T^{-1/2}$ plot at various fluences. The inset shows resistance ($R$) versus temperature ($T$) plot. b) Optical absorption spectra of ion exchanged and irradiated samples at various fluences. (Ref. [51]@2001,Copyright©*Elsevier Science B.V.; Applied for permission*)

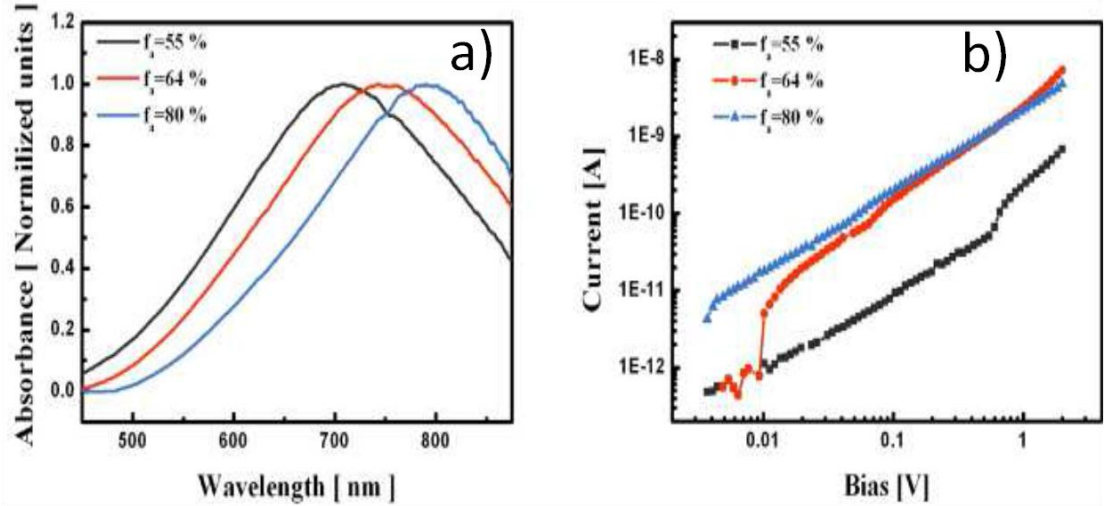

**Fig. 5.** (a) LSPR measurement results from samples with a surface coverage, $f_a$ of 55%, 64%, and 80%; each samples has a different resonance peak (55%: 706 nm; 64%: 743 nm; 80%: 789 nm). (b) I-V characteristics of samples with a surface coverage of 55%, 64%, and 80%; the plot of $f_a$= 64% shows a jump of about one order of magnitude (for the percolation threshold). (Ref. [41]@2010,Copyright©*American Optical Society; Applied for permission*)

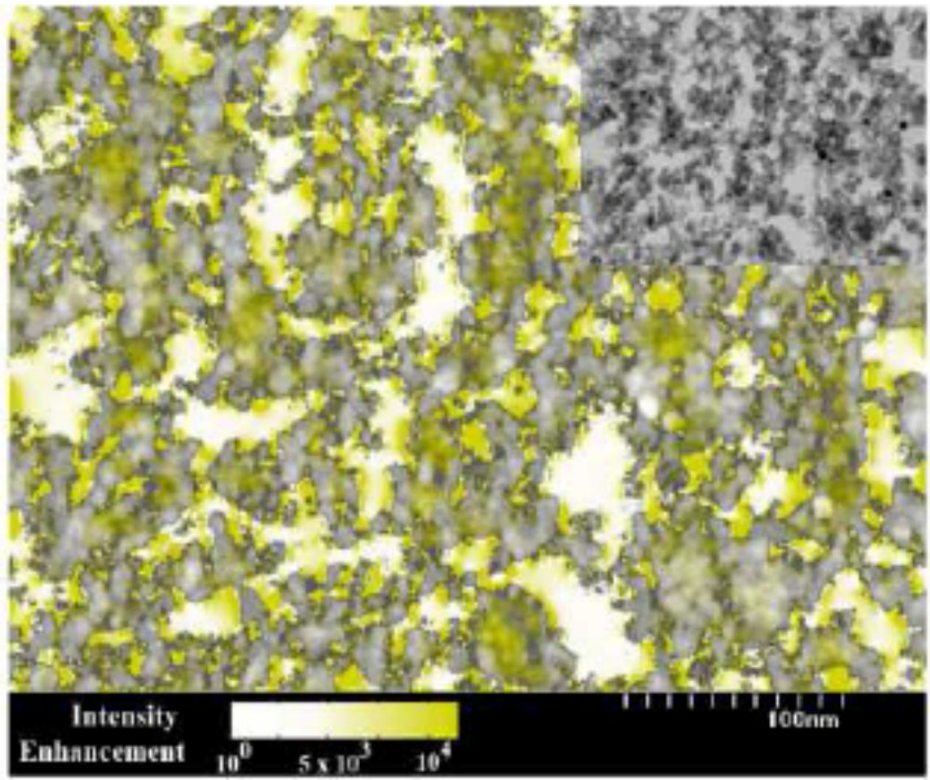

**Fig. 6.** Calculated electric field intensity enhancement in the plane of the deposited nanoparticles (64% coverage sample). The local field intensity enhancement is depicted in the TEM image (inset) using the linear color bar. (Ref. [41]@2010,Copyright©*American Optical Society; Applied for permission*)

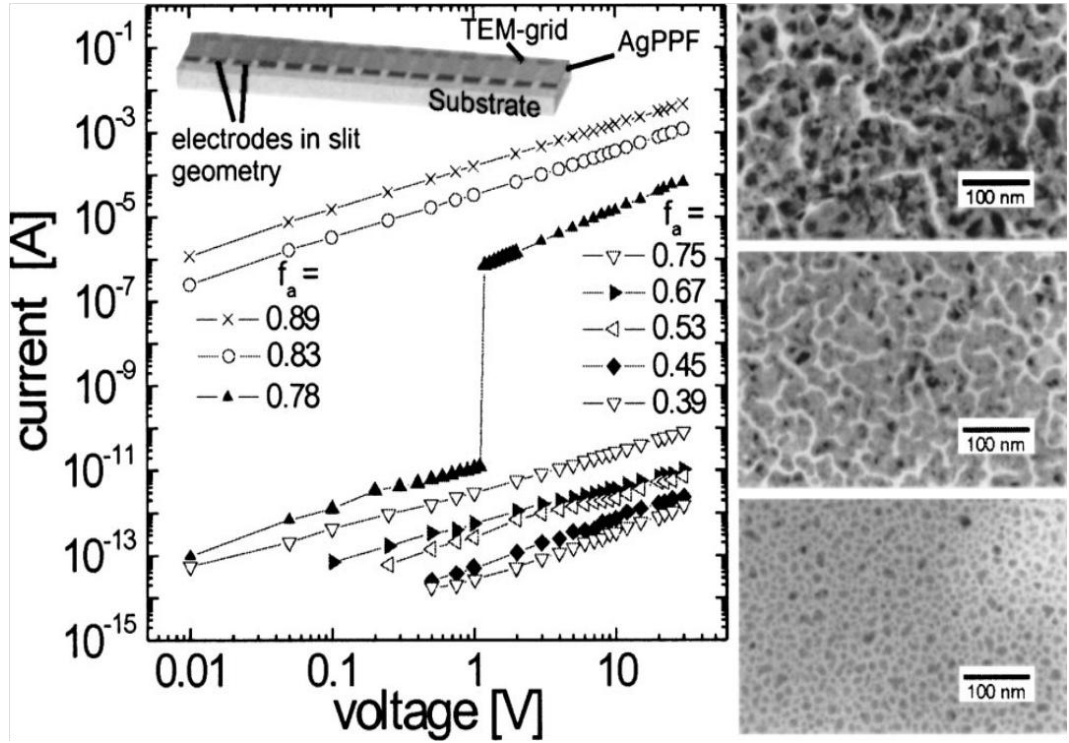

**Fig. 7.** Left: $I-V$ characteristics of a plasma polymer film containing Ag nanoparticles (AgPPF) with different area filling factors $f_a$. Top left: Sample setup: Coplanar electrode arrangement (slit dimensions 500 μm×6 nm); placement of TEM grids is matched to film positions (slits) for the electrical measurements. Right: TEM micrographs of different nanostructural types (dark: metal; light: plasma polymer). (Ref. [42]@2003,Copyright©*American Institute of Physics; Applied for permission*)

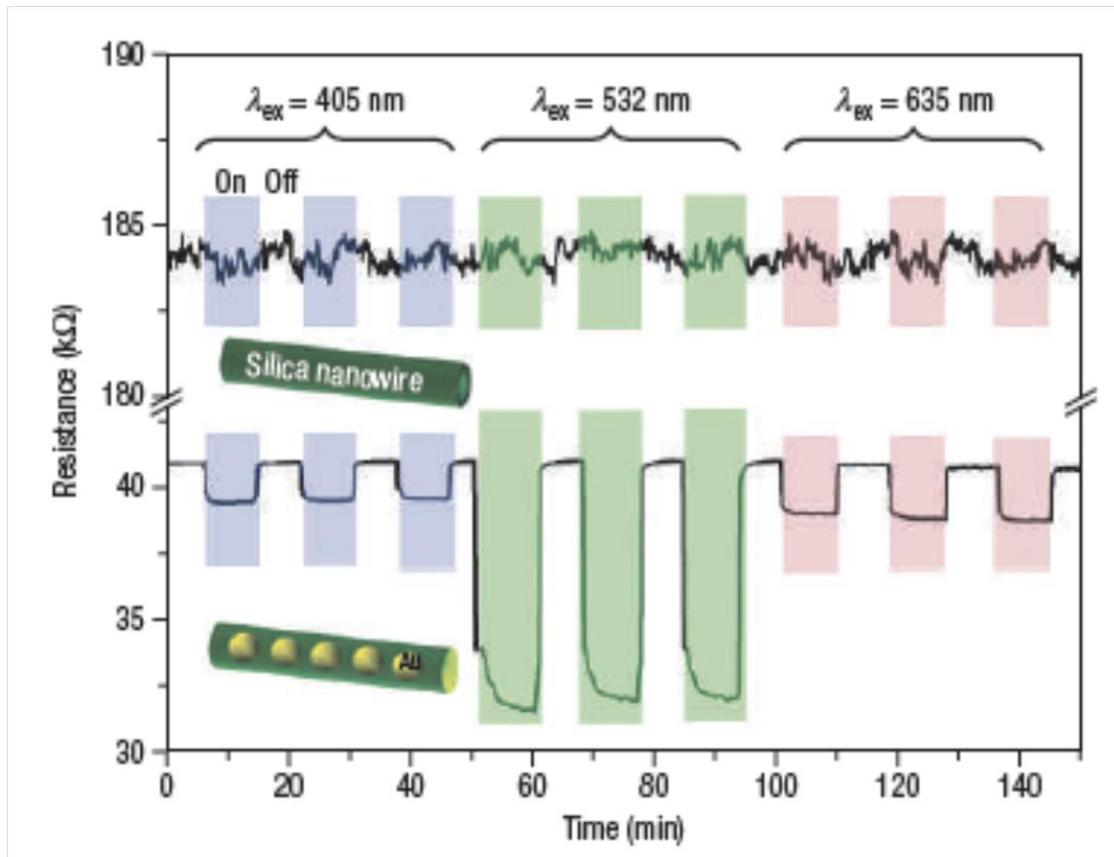

**Fig. 8.** Photoresponse measurements. The room-temperature resistance response as a function of time to light illumination for plain silica nanowires (upper part) and gold nanopeapodded silica nanowires (lower part). Shaded (pink, excitation wavelength $\lambda_{ex}$ =635 nm; green, $\lambda_{ex}$ =532 nm; purple, $\lambda_{ex}$ =405 nm) and unshaded regions mark the light-on and light-off periods, respectively. (Ref. [33]@2006,Copyright©*Nature Publication Group; Applied for permission*)

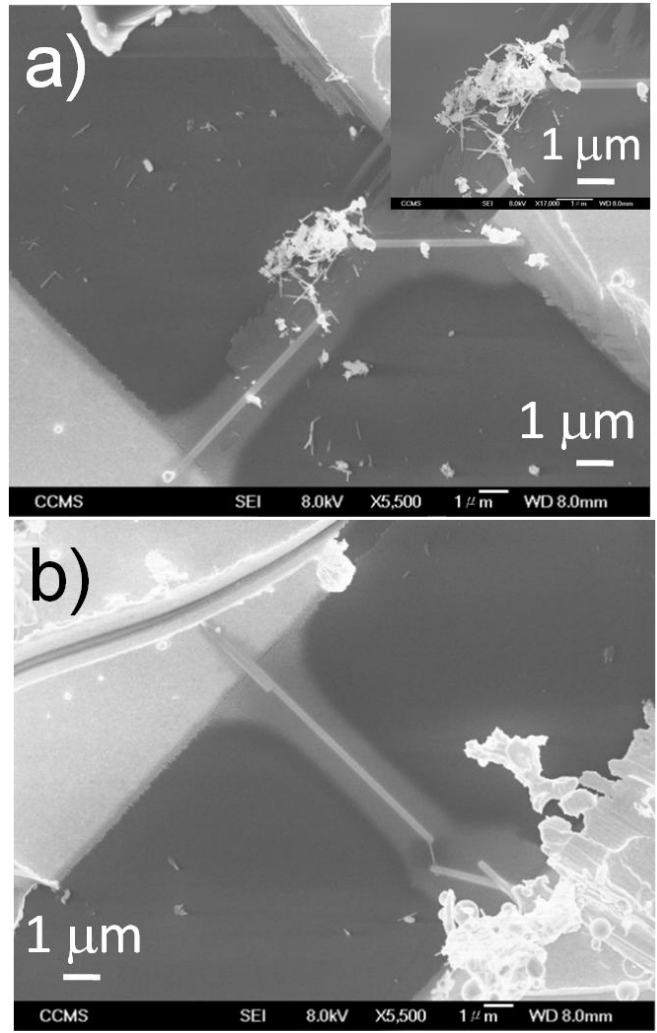

**Fig. 9.** Focused ion beam assisted Pt contact electrodes in (a) ensemble (b) single GaN nanowire system. (Ref. [34]@2014,Copyright©*Springer; Applied for permission*)

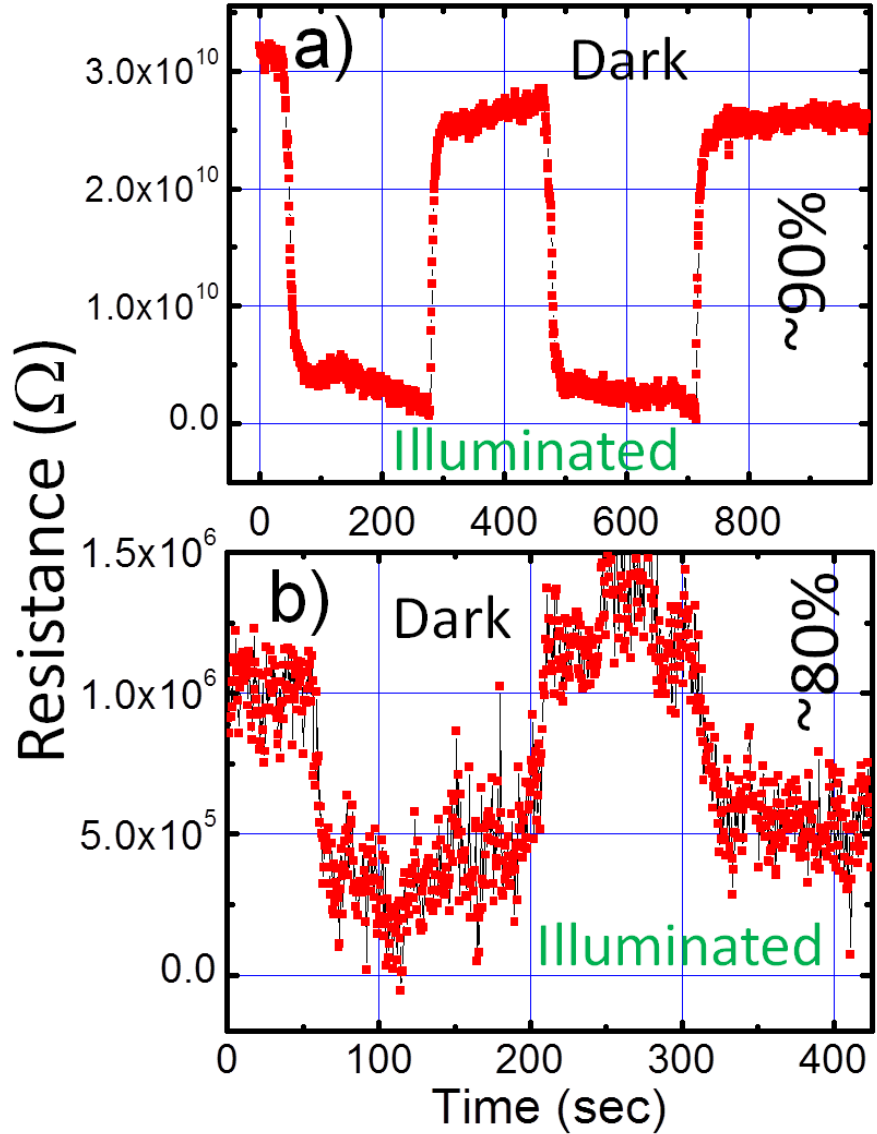

**Fig. 10.** Photoresponse in a) ensemble b) single GaN nanowire samples with periodical dark and 532 nm illumination conditions. (Ref. [34]@2014,Copyright©*Springer; Applied for permission*)

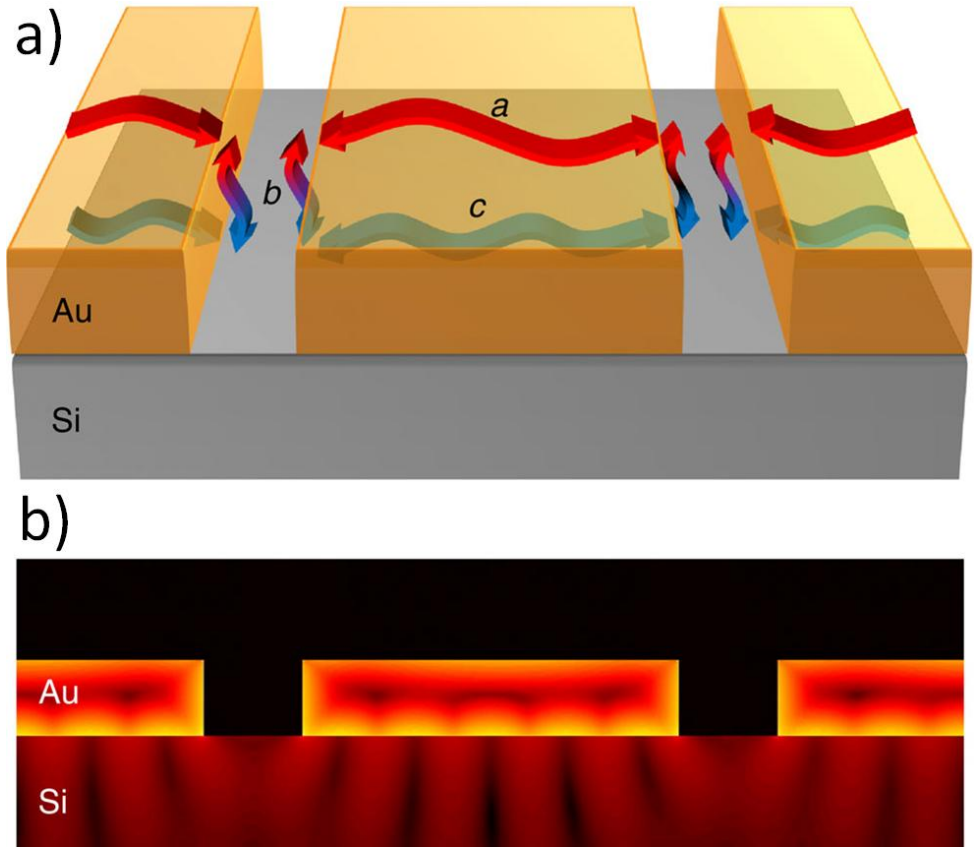

**Fig. 11.** Propagation of surface plasmons on gratings. (a) There are three forms of surface plasmon polaritons (SPPs) around each gold pitch: *a*, SPPs oscillating at the top surface of gratings; *b*, SPPs oscillating though the slits and; *c*, SPPs oscillating at the bottom surface of gratings at the Ti–Si Schottky interface. (b) Plasmonic heat absorption calculated by FDTD (shown with a logarithmic scale for clarity). Most of the hot electron generation occurs at the bottom surface of the gold layer. ([Ref. [55]@2013, Copyright©*Nature Publication Group; Applied for permission*)

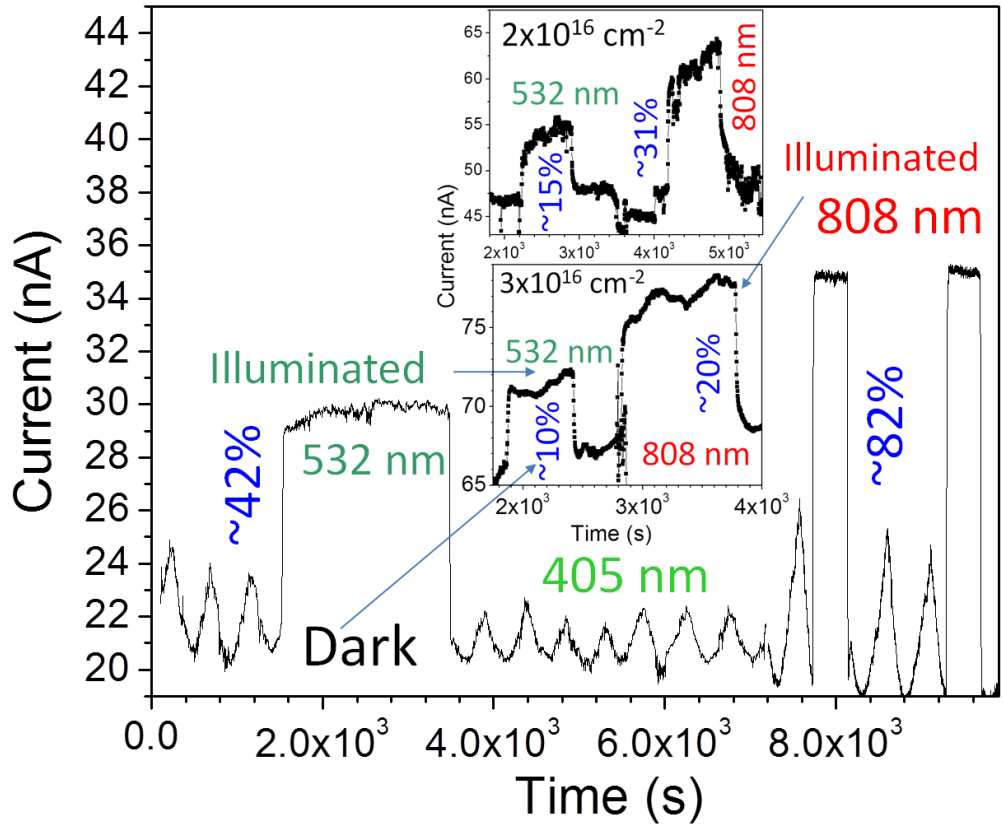

**Fig. 12.** Typical photoresponse of Au-Ag embedded in soda glass sample grown with $Au^+$ implantation at a fluence of $5\times10^{16}$ ions.cm$^{-2}$ with periodical dark and illumination to different laser wavelengths. Inset shows similar photoresponse of samples grown with $Au^+$ implantation at fluences of $2\times10^{16}$ and $3\times10^{16}$ ions.cm$^{-2}$. The studies show a double amount of change in current for 808 nm excitation than that observed for 532 nm exposure. (Ref. [40]@2015,Copyright©*American Institute of Physics; Applied for permission*)

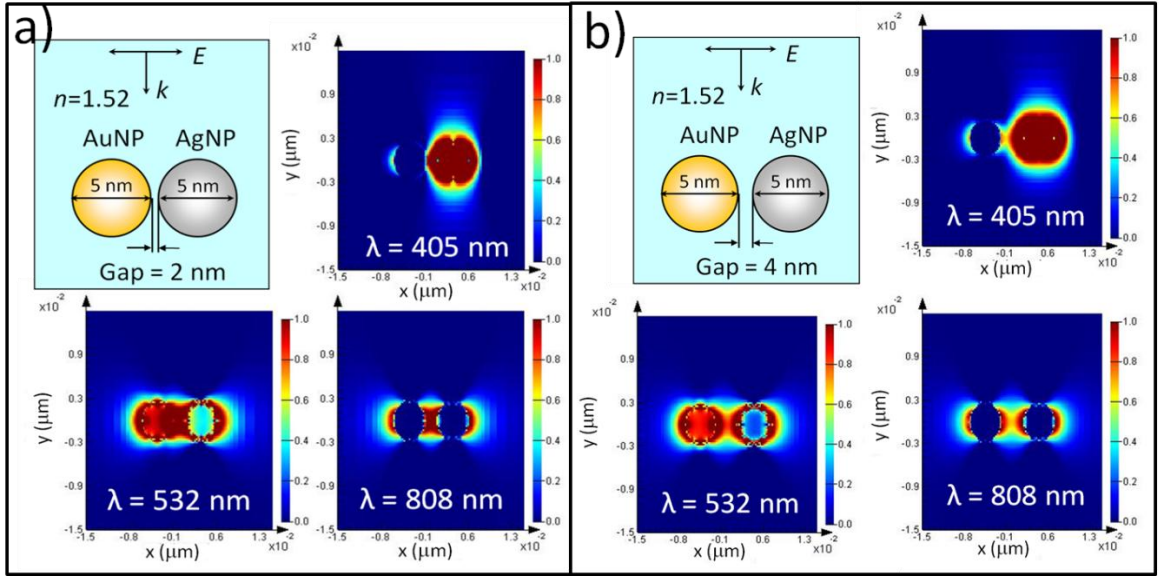

**Fig. 13.** Comparison of the electric field intensity enhancement contours for the interaction of electromagnetic radiation of different wavelengths with 5 nm Au and Ag nanoparticles separated by a) 2 nm and b) 4 nm in a medium with refractive index *n*=1.52. Here *k* is the electro-magnetic wave propagation vector, *E* is the electrical field vector and intensity bars indicate $|E|^2$. (Ref. [40]@2015,Copyright©*American Institute of Physics; Applied for permission*)